# Managing Requirement Elicitation Issues Using Step-Wise Refinement Model

Nikita Nahar G
PG Student,
Dept of Computer Science
MOP Vaishnav College for
Women, Chennai, India
gniki93@gmail.com

Pujita K wora
PG Student,
Dept of Computer Science
MOP Vaishnav College for
Women, Chennai, India
pujitawora2012@gmail.com

Sakthi Kumaresh
Asso. Prof,
Dept of Computer Science
MOP Vaishnav College for
Women, Chennai, India.
sakthimegha@yahoo.co.in

*Abstract*— **Requirements engineering is the fundamental aspect of Software Process development. It is modularized into various stages of elicitation, Analysis, Validation and Documentation. The process of building software starts with requirement elicitation. This phase determines what problem needs to be solved or what software has to be developed. Requirements are constraints or demands which are described by the customers and must be met within a specific time period. While, Elicitation is all about gathering necessities and requirements from the end-user's to recognize what a software system should consist of, acquisition of requirements is done via an interaction between the analysts and the stakeholders. If the need of the stakeholders doesn't meet then the quality software will not be produced.**

**If an error occurs in Requirement Elicitation stage, then it will be carried on to the further stages which will deteriorate the whole product. So, it is very important to fix the errors at the right stage. This phase acts as a foundation for developing a project and the role of the customers is most essential. Many challenges arise during this process; if faulty requirements are encountered during this stage then they will produce ripple effect on further software development process.**

**Thus, realizing the importance of Requirements Elicitation Stage, there is a need for more commendable model through which requirements can be elicited in a much simpler manner. In this study, a Step-wise Refinement model is proposed to elicit requirements in a more effective manner. According to this model, requirements are elicited after doing domain and stakeholder analysis which is then analyzed with the help of customers and finally a graphics prototype is developed for customer's approval. The most important asset of this model is that the stakeholder's will have a pre-hand notion about how the software system will look like which will in turn reduce the majority of defects at the later stage of the project.**

*Keywords- Requirements, Stakeholders, Software, Quality, Elicitation.*

## I. INTRODUCTION

Requirement Elicitation is the process of gathering, collecting, acquiring and detailing the needs of the customers to build up a quality product. This is the earliest and most challenging phase in building a project and acts as a root for developing a project. The main goal of this stage is to satisfy the client's needs and not to overrun the cost and time of the project.

Requirements can be classified into two types Known and Unknown requirements. Known requirements are those which the customer knows and want to implement [1].

Unknown requirements are those which the stakeholders have forgotten or they may not be aware of it [1]. This notifies us that requirements must be collected in such a manner that known and unknown both are included while identifying their specifications.

Known or Unknown requirements may be functional or non-functional [1]. The Functional Requirement documents the operations and activities that a system must be able to perform [2]. Non functional requirements are the properties that your product must have [3]. These prerequisites collected must be correct, complete and unambiguous. Only then it would be easier for the designers to transform it into meaningful design descriptions. Bad requirements may lead to budget overrun, cost reworks, poor quality product and dissatisfaction of customers and a project failure.

Despite of many years of research, requirements are still ill-defined and are added even after its elicitation phase. This calls for a need to develop a model. It is quoted "Requirements cannot be observed or asked for from the users but have to be created together with all the stakeholders" [4]. This means that Elicitation stage can succeed only through an effective client-analyst partnership, such that there is no misunderstanding between them.

Role of the Stakeholders is the utmost important in developing a Software system because only with the help of the customers, needs of the software can be captured, understood and validated. The requirements should be collected from those stakeholders who have better domain knowledge. So that rich information may be provided. Requirements can be acquired through asking questions, writing down the answers, conducting interviews and by appropriate elicitation techniques. These techniques are generally used to assist analysts in identifying the user needs.

There are various issues confronted during this process. As the customers and developers emerge from different backgrounds, misunderstanding between them can arise.

Changing nature of requirements, inadequate communication, problem of scope, incomplete requirements, ambiguous requirements, wrong selection of stake holders, inappropriate selection techniques, conflicting requirements are some of the problems encountered during this phase [1]. Indeed, some research has concluded that systems failure can be traced back to poor requirements elicitation in up to "90% of large software projects" [5].

Through this paper we have proposed some possible interventions where a Step-wise Refinement model has been developed to overcome the issues which arises during this phase and provides a graphics prototype.

## II. CHALLENGES FACED DURING REQUIREMENT ELICITATION PHASE

A requirement is an expression of desired behavior [6]. A requirement deals with objects or entities, the states they can be in, and the functions that are performed to change states or object characteristics [6]. Requirement elicitation is the most vital and crucial phase of software engineering, because errors occurring during this stage propagate throughout the development phases and are hard to repair later. The mindset of many requirement engineers is that, collecting the requirements from the customers is an easy task and doesn't require much effort, but in reality it is contrary. There are various issues that may be confronted while gathering the requirements from the end- users. We have examined few of these issues in this section.

- Requirements acquired can be contradicting and conflicting, which means that different stakeholders may provide different requirements and opinions at different time, which would be difficult to implement. This may lead to poor specifications and the cancellation of system's development.

- Specifications of bad requirements by the customers, is a major defect encountered. End users usually specify those requirements that are not at all necessary. Eliminating these requirements results into a time consuming process.

- Ambiguous requirements should be avoided. Specifications imparted from the customers have different meaning and results in misunderstanding. This issue is highly sensitive and should be taken care of.

- The next challenge faced during this stage is lack of clarity of the requirements. Unclear and obscure requirements when collected will complicate the developer's task. The developer might not be able to understand what user's actual need is and what has to be implemented into the software. Such requirements have to be gathered, which can be easily transformed into design descriptions and can be coded by the developer.

- Developer may develop the project without being aware of what the end product should consist of. There may be problems in understanding from the developer's side. Although definite requirements are provided by the customers, clarity of the same should be maintained on the developer's side. The requirements may not be well managed. So, it is very important for the developers to understand the background details of the customer's requirements.

- Various techniques are applied to collect the actual essentials from the customers. It is prominent that the requirements acquired should be written or documented accurately as it would be easier for the developer to implement it. These documents act as an agreement between the clients and the requirement engineers, describing what the user need is and what the system should actually consist of. Errors occurred during this level leads to cost of rework and time delay.

- Requirements specified by the customers are volatile in nature. Henceforth the needs of the customer change intensely as per the market conditions. If customer interaction is not taken into consideration then end product will not be according to the customer needs. Thus satisfaction of customer will not be attained and failure of the product intensity will increase.

- Communication gap between the two parties is a major area of concern. To acquire information from the customers, developers are ought to interact with them. But, the stakeholders and analysts both belong to different fields. The customers talk in general language according to their business perspective while the analysts speak in technical term which leads to misinterpretation.

- The other issue related to communication challenge is many customers don't speak up for the requirements to be implemented in the software or are too shy to tell the developers. If this problem persists, then eliciting requirements becomes difficult and user's view of the system cannot be captured and implemented.

- The most critical issue is nonusage of the existing information available with the software developing company that has developed similar software previously. Thus it will result in increased cost and time, in eliciting the same set of requirements.

Hence the focus of this paper is to highlight the issues concerned with requirements elicitation and to overcome them using Step-wise Refinement Model. Poor requirements are one of the biggest reasons for failure of projects [7]. By reforming the mode of capturing the requirements, we can reduce the total number of defects resulting in enhanced system requirements.

### III. STEP-WISE REFINEMENT MODEL

Step-wise Refinement Model is a model devised capable of overcoming the hurdles faced during gathering of requirements. The model is named so, because after each stage the requirements are refined and are projected to the customers using graphic prototype for evaluation. If the customers are not satisfied with the design of the prototype, then modifications are made as per the feedback given by the customers. Graphic prototype is the highlight of this model as the customers are given an opportunity to view a prototype of their software which will build the confidence of developers. On the other hand, it will serve as a guide to the developer for initiating the software development process.

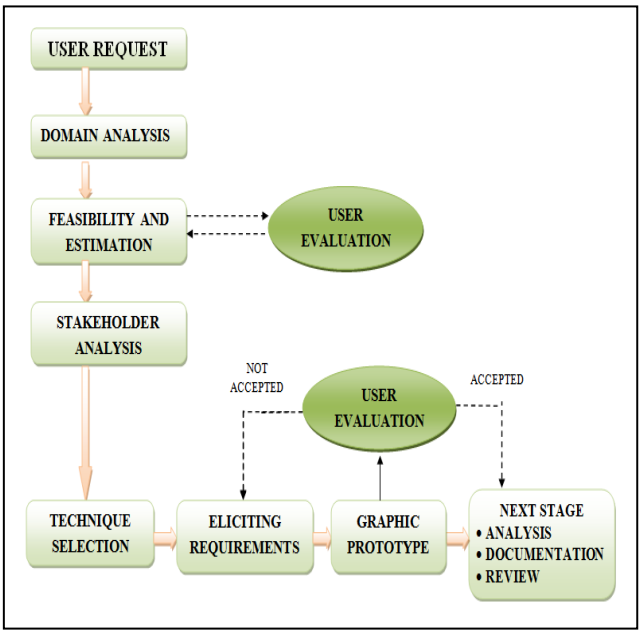

Figure 1 Step-wise Refinement Model.

Figure1 depicts the Step-wise Refinement model which consists of 8 stages starting from user request phase to the review phase. A brief explanation of each stage in step-wise refinement model is done in the following lines:

*1) User request:* It is the first phase in the process of Requirement Elicitation. In this phase, a customer demanding software to be built on the basis of its specifications approaches a software developing company. The customer is expected to mention the type of software it requires to satisfy its needs. Therefore, it is the stage at which a link is established between the developer and the customer.

*2) Domain analysis*: It is the duty of the developer to determine the background information based on the type of software requested by the customer. Therefore, the developer in this stage gathers information which will help him in understanding the problem stated by the user and in making decisions for further stages. The Requirement's Engineer also verifies whether the software or modules related to the system to be developed exits so that it can be reused. This would guarantee the development of quality software.

*3) Feasibility and estimation:* A feasibility study is an evaluation and analysis of the potential of the proposed project which is based on extensive investigation and research to support the process of decision making [8]. A feasibility study is done on the basis of three factors such as technical, legal and economical feasibility.

- Technical feasibility: It focuses on the availability of technical resources in the organization. It also verifies whether it meets the user needs or not.

- Legal feasibility: It verifies whether the product to be developed does not conflicts with legal requirements

- Economical feasibility: It assess whether the organization will be economically benefitted or not.

If all the three factors are in favor of the software, then no change is requested and it is carried on to the next stage or else change request will be made. In the latter case, the changes are to be evaluated by the customers and based on which decisions on approval is made. Hence user interaction plays an important role in this phase.

*4) Stakeholder analysis:* Stakeholders refers to the individuals who are affected by the software, either directly or indirectly. Analysis of stakeholders refers to the selection of the stakeholders who have strong domain knowledge about the software and its functioning and also have experience in using the software. Hence the role of stakeholders is of utmost important in developing a software system. The inputs provided by them are valuable to the developers as it gives a clear vision of their expectations. Selection of the stakeholders varies from organization to organization.

*5) Selection of technique:* The next phase in the model is the selection of an appropriate technique to collect the requirements from the stakeholders. The stakeholder's analysis phase may not provide in-depth information about all the requirements. Therefore, to ascertain the requirements in detail, various techniques such as Interview, Brainstorming, FAST, and Quality Function Deployment and so on can be used. The developers can use techniques according to the type of software requested to be built. A single technique or a series of techniques can be used to develop a project. These techniques are generally used to assist analysts in identifying the user needs. Failure in selection of appropriate techniques may increase the cost and time of development.

*6) Eliciting requirements*: This is the most important phase of the Step-wise Refinement model. Here, with the help of various stakeholders and techniques selected, elements are collected. Different Stakeholders provides various requirements about the system to be built. The stakeholders provide details regarding functionality, performance and hardware constraints of the system. Both, the customers and the developers interact to elicit the essentials. If this phase is not carried out in an appropriate manner then it might result in bad Requirement Specification. The requirements collected are recorded into graphics prototype.

*7) Graphic prototype:* This is the most curious phase of requirements elicitation. At this stage the customers are shown a design of the software. The prototype is made using any graphic software which the designers excel in or the company has. This prototype design serves as the basis on which the development process begins. Hence, this phase is the primary stage for further process. If the customer approves the graphic prototype, then it goes to next stage. Otherwise, the requirements have to be properly elicited again. Henceforth changes are made in the prototype according to the given new requirements and thus evaluated by the user. So the process goes on till the customer approves it and carried on with the next stage.

*8) Next stage:* The elicitation stage ends with the evaluation of graphic prototype by the stakeholders. After requirements elicitations there are three more steps to be followed to complete the requirement engineering process. They are:

- Requirement analysis: In this, requirements that are gathered in elicitation stage are analyzed for correctness and completeness

- Requirement documentation: Once the requirements are analyzed they are documented which is nothing but 'software requirement specification' is prepared.

- Requirement review: In this phase, the requirements documented are reviewed so that software requirements specification's quality is increased and it can be passed onto other development stages.

Thus these are the steps which should be pursued to gain a successful requirement elicitation. By the means of this model, we can prove that user involvement in each phase of the development of the software is of paramount importance as they ensure the development of quality-oriented software.

## IV. CASE STUDY

*HOSPITAL MANAGEMENT SYSTEM*

Hospital Management System is a software product suite designed to improve the quality and management of clinical care and hospital health care management in the areas of clinical process analysis and activity-based costing [9]. It provides extensive solution to various healthcare industries that includes multispecialty or general hospital. The operations included in the software are patients, doctors, ambulatory, nursing, ward and so on. . The success of Hospital management system is possible if the requirements of the end-users are satisfied.

'Himalaya hospital' is a multispecialty hospital that has implemented hospital management system (HMS). The HMS modules for this hospital include patient, appointment scheduling, admission, medical documentation and services, ambulatory, doctor, nursing, operations room, employee, ward management, laboratories and radiology. The reason for projecting 'Himalaya hospital' is because; it is suffering from various defects while using hospital management system. Few of the defects are mentioned below, they are:

- The main issue encountered by the Himalaya hospital is, there is no such module which will update the supplier when the medicine stock is empty, and hence new stock can be sent by the supplier. For example, if there is a sudden rush for particular medicines, there is no risk of the shelf becoming empty, as the particular medicine supplier would get a constant computer-generated update by the computer system on the medicine being sold.
- Packages and health plans was not included in the modules.
- The module 'employee feedback' exists in the software which is not at all required. This is an extra module developed by the software company.
- The other important criterion is about the tax rate of medicine which has increased but it cannot be changed. This leads to the loss of hospital.
- Instead of 'out-patient (patient who visits the hospital for treatment)' which has to be included in the patient module was included in admission module (update the details of admitted patient). This was due to the misunderstanding between the customer and the developers' side.

The above project was confronted with various defects while using the HMS software. The reason behind it is the usage of traditional elicitation process to develop the software. Whereas, compared to the previous project, the defects encountered in Resort Management Software were very minor as the Step-wise Refinement model was exerted to develop a quality product. This is achieved, as it includes customer's evaluation in each phase. The Step-wise Refinement model was implemented in the Resort Management System as shown below:

*RESORT MANAGEMENT SYSTEM*

The research paper projects the case study of Resort Management system to show that the Step-wise Refinement Model has been implemented to overcome the challenges faced during the Requirement Elicitation process. Resort Management system is a project, developed to enhance the automation of Resort Management System. This project aims to administer various details of the customers, employees, departments and other hotel details where only authenticated users can work with these details.
Step-wise Refinement Model has been implemented to gather the requirements from the users for this specific software. A step-by-step explanation of these stages is given with respect to the Resort Management System (RMS).

*1) User Request:* The Client fixes up a meeting for placing a request to a software developing company to build RMS. The customer briefs the company about 'what' the software is.

*2) Domain Analysis:* The Requirements Engineer performs a background study about the RMS to understand the problem domain and to make good decisions during the requirement elicitation. There is a check within the software developing company whether any software or modules related to Hotel or Resort Management has been developed.

*3) Feasibility Study and Estimation:* Engineers determines the budget and the period of time interval to build RMS. Analyst's ensures that the customer's company has particular compatible operating system and technical tools without which RMS cannot be executed. According to these constraints developing company places a negotiation deal. The client's company approves with it and developers start the work to proceed with further stages of RMS.

*4) Stakeholder Analysis:* The customers are selected as the stakeholders to gather information about 'what' has to be implemented. In defining RMS, Managing director of the Client's company, Head of the Accounts & Finance, Food Service departments, the Receptionist and 1 staff member from each of the department are selected as the stakeholders. As they have rich knowledge in their particular domain.

*5) Selection of Techniques:* Developer's has to determine the techniques through which they can acquire information from various stakeholders. Structured Interviews, Brainstorming, Facilitated Application Specification technique (FAST) were selected to gather information. FAST technique was used to bridge the gap between the Developers and the customers to elicit those requirements of which, the user cannot think of its implementation.

*6) Eliciting Requirements:* With the help of various selected stakeholders and techniques modules for RMS are identified. Six major modules of user-interface are identified as Check-in, Check-out, Advance Booking, Employee Master, Room Master, Food Service and in-depth information about these modules are collected.

*7) Graphics Prototype:* The Elements elicited are transformed into graphical components which describe how the system will look like. If the client doesn't agree with this prototype, re-gathering of requirements will take place and changes will be incorporated within the prototype.

*8) Next Stage:* After the evaluation of Graphic prototype by the customer, analysis of the requirements gathered is done. Pre-requisites collected are documented and finally reviewed by the Client and software developing company to carry out further software development process.

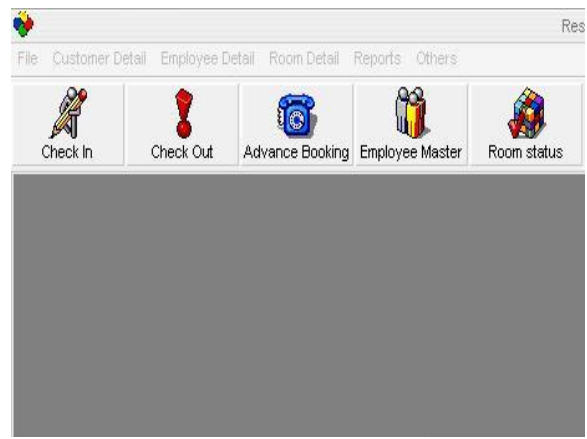

Figure 2 Modules of Resort management system shown with the help of graphic prototype.

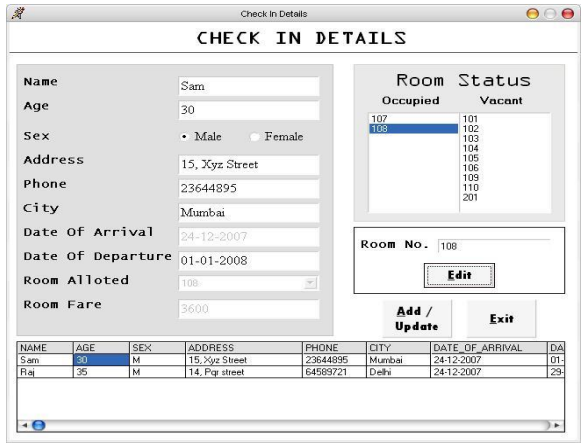

Figure 3 Resort management system's check-in details.

We have tried to overcome many issues which were faced during the elicitation phase. The Step-wise Refinement model has helped us to eradicate few of the major issues. They are:

• Usually feasibility study is carried out only after the requirements are elicited. If the client has problems with the feasibility study conducted, he will terminate the contract which will effect directly on the engineers efforts of eliciting the requirements. This challenge has been overcome by stepwise-refinement model where feasibility study is conducted after the domain analysis.

• In the domain analysis, the developer verifies whether a particular software or module related to the problem is available. If the information related to the developing software is available then these modules are brought into use. Through this, the defect of non-usage of existing information is rectified.

• Developers before eliciting the requirements performs domain analysis in which background details of the software to be developed is analyzed. After performing this step, the engineer gains a clear idea about the software to be developed. The problem in understanding the requirements by the developers is solved here.

• Irrelevant requirements provided by the stakeholders can be removed during the evaluation of graphics prototype by the customer.

• User interaction persists while functioning out every level in step-wise refinement model which resolves the issue of misunderstanding between them.

• Generally, after eliciting the requirements they are documented and analyzed by the customers. The client will be unable to understand what developer actually has collected through the testament. It is easy for the customer to figure out through a real time prototype. Incorporation of Graphics prototype helps the customer to know what will be implemented.

• During this elicitation process, if incorrect stakeholders are selected then it may result in acquisition of bad requirements. To solve this issue, selection of stakeholders is done on the basis of the software to be built and the people who use it. In RMS receptionist is selected as the stakeholder as he/she is the one who interacts with the software on day to day basis.

• Incorrect use of elicitation technique results in time consuming process. When changes are to be done in existing software then through the help of structured interview technique requirements can be elicited. If FAST technique is followed, it will lead into a time consuming process. So before the gathering of requirements start, thorough analysis of the techniques has to be done

Table 1- Comparison between Hospital Management System and Resort Management System

| Project name | Hospital Management System | Resort Management System |
|---|---|---|
| Project description | HMS was encountered with many defects. Because many errors occurred while eliciting the requirements and were carried onto further software development process. | RMS overcomes many issues due to Step-wise Refinement model. As reason being, this model involves user interaction in each phase. |
| KLOC(size) | 2000 | 1800 |
| EFFORT (person months) | 20 | 18 |
| No of defects found | 370 | 275 |
| Defect density | 0.18 | 0.15 |

Formulae to calculate defect density:
 Defect density=No of Defects/KLOC [10]          (1)

Table 1 shows the two projects HMS and RMS. It is noted that the effort required and defect density of RMS is less when compared to HMS. This is depicted graphically through figures 4 and 5.

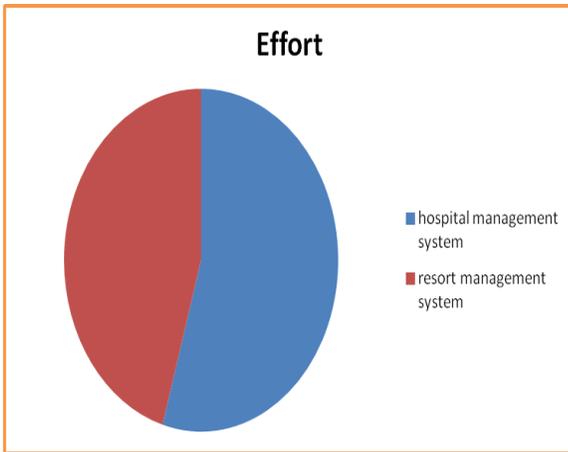
Figure 4 Chart Showing Effort Applied In Person-Months.

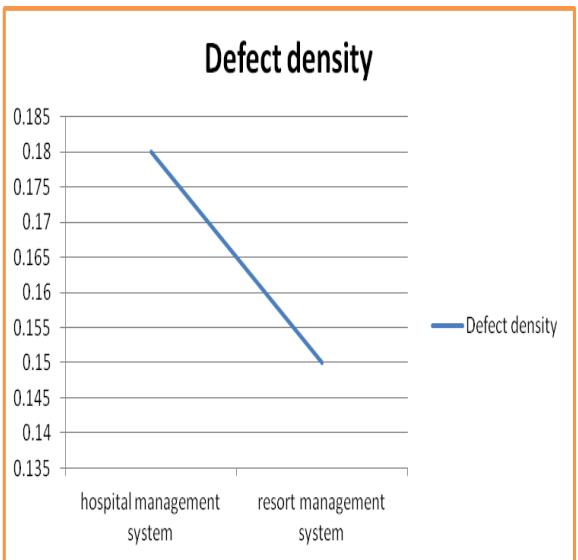
Figure 5 Graph Showing Defect density.

## V. CONCLUSION

Requirement elicitation encounters various challenges, while gathering the requirements from the stakeholders. This paper discusses how to overcome these challenges using Step-wise Refinement model. Step-wise Refinement model includes user interaction in each phase. When compared with traditional elicitation models, the defect density and effort applied in person months is less for step wise refinement model. The project "Resort management system" is build using SDLC model but the requirements are gathered using the Step-wise Refinement model and hence the base of the software (requirements gathering) is done in an efficient manner. Thus the further development of the software is carried out smoothly with less number of defects. The defect rate and the defect density is comparatively less when compared to the other project in which the requirements are gathered using the basic elicitation techniques. The major advantage of using the step wise refinement model for requirements elicitation is the "Graphic prototype phase" where a graphic component is formed that resembles the system. The client can have a feel of it and can easily make out problems if any. As requirements forms that base of the software, it is very important that there are no problems in gathering requirements. Thus this research paper projects a model to elicit requirements in a step by step manner more efficiently.